\documentclass[journal=jpccck,manuscript=article]{achemso}

\usepackage[version=3]{mhchem} 

\author{I. Perez}
\email{cooguion@yahoo.com}
\affiliation{Department of Physics and Engineering Physics, University of Saskatchewan, 116 Science Place, S7N 5E2, Saskatoon SK Canada}

\author{J. A. Mcleod}
\affiliation{College of Nano Science and Technology, Soochow University, 199 Ren-Ai Rd., Suzhou Industrial Park, Suzhou, Jiangsu, 215123, China.}
\altaffiliation{Department of Physics and Engineering Physics, University of Saskatchewan, Saskatoon SK Canada, 116 Science Place, S7N 5E2}
\author{R. J. Green}
\affiliation{Department of Physics and Astronomy, University of British Columbia, 6224 Agricultural Road, Vancouver, British Columbia V6T 1Z1, Canada.}
\altaffiliation{Department of Physics and Engineering Physics, University of Saskatchewan, Saskatoon SK Canada, 116 Science Place, S7N 5E2}

\author{R. Escamilla}
\author{V. Ortiz}
\affiliation{Instituto de Investigaciones en Materiales, Universidad Nacional Aut\'onoma de M\'exico,  M\'exico D.F.04510, Mexico}
\author{A. Moewes}
\affiliation{Department of Physics and Engineering Physics, University of Saskatchewan, 116 Science Place, S7N 5E2, Saskatoon SK Canada}
\title[Electronic Structure of $\textrm{Fe}\textrm{Se}_{1-x}\textrm{Te}_x$]
  {Electronic Structure of $\textrm{Fe}\textrm{Se}_{1-x}\textrm{Te}_x$ Studied by X-ray Spectroscopy and Density Functional Theory}

\abbreviations{XRD,DFT,RIXS,XES, RR, NRR}

\SectionNumbersOn
\begin{document}
\maketitle
\begin{tocentry}

\includegraphics[width=6cm]{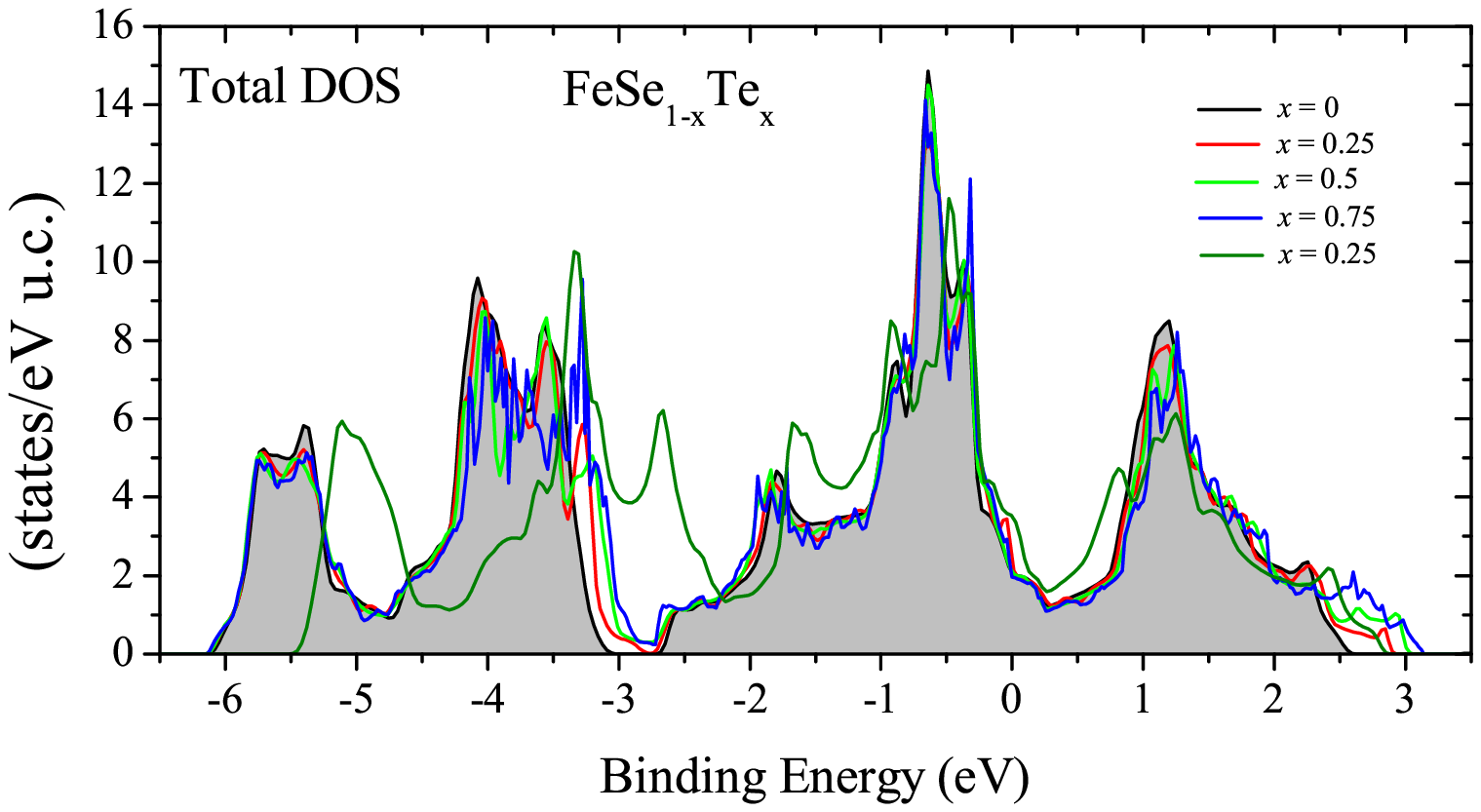}

Total density of states for the $\textrm{Fe}\textrm{Se}_{1-x}\textrm{Te}_x$ system
\end{tocentry}

\begin{abstract}
We study the electronic properties of the $\textrm{Fe}\textrm{Se}_{1-x}\textrm{Te}_x$ system ($x=0$, 0.25, 0.5, 0.75, and 1) from the perspective of X-ray spectroscopy and density functional theory (DFT). The analysis performed on the density of states reveals marked differences in the distribution of the $5p$ states of Te for $x>0$. We think that this finding can be associated with the fact that superconductivity is suppressed in FeTe. Moreover, using resonant inelastic X-ray scattering, we estimate the spin state of our system which can be correlated to the magnetic order. We find that the spin state of the $\textrm{Fe}\textrm{Se}_{1-x}\textrm{Te}_x$ system fluctuates, as a function of $x$, between $S=0$ and $S=2$ with Fe in FeSe in the highest spin state. Finally, our DFT calculations nicely reproduce the X-ray emission spectra performed at the Fe $L$-edge (which probe the occupied states) and suggest that the $\textrm{Fe}\textrm{Se}_{1-x}\textrm{Te}_x$ system can be considered at most as a moderately correlated system. 
\end{abstract}

\section{Introduction}
The discovery of superconductivity at 28 K in the system LaFeAs($\textrm{O}_{1-x}\textrm{F}_x$) by Kamihara et al. \cite{kamihara}  triggered a great deal of research on the study of the physical properties of iron-based layered superconductors. The situation with these materials is quite comparable to that of the cuprates during the 1980s. Soon after, superconductivity was reported at critical temperatures ($T_c$) of 38 K, 18 K, and 8 K in the systems $\textrm{BaFe}_2\textrm{As}_2$, $\textrm{Li}_{1-\delta}\textrm{FeAs}$, and $\textrm{FeSe}$, respectively \cite{rotter,wang5,hsu}. In these investigations it was found that all iron-pnictide superconductors possess a two-dimensional Fe-pnictogen layer with a tetragonal structure at room temperature. Therefore, their physical properties are considered to be highly two-dimensional, similar to cuprates. Despite great efforts realized in previous years, the understanding of the mechanisms responsible for the emergence of superconductivity still remains unclear. From the theoretical point of view, binary or ternary systems such as $\textrm{FeSe}$ and $\textrm{Li}\textrm{FeAs}$ have been considered more suitable models for the study of superconducting and electronic properties because they are structurally simple. Band structure calculations have shown that $\textrm{FeSe}$ and the pnictogen compounds have similar Fermi-surface structures \cite{ma} implying that $\textrm{FeSe}$ can significantly contribute to elucidating the origin of superconductivity. Moreover, $\textrm{FeSe}$ has a tetragonal $\alpha$-$\textrm{PbO}$-type structure at room temperature composed of planar layers of $\textrm{Fe}_2\textrm{Se}_2$ which are similar to the layers in the pnictogens. Although $\textrm{Li}\textrm{FeAs}$ has a higher $T_c$ than $\textrm{FeSe}$, it possesses one additional planar layer of Li, making it more complicated. Furthermore, $\textrm{FeSe}$ has no toxic compounds, and it is, therefore, more desirable for commercial applications. 

Extensive work on the synthesis and characterization of $\textrm{FeSe}$ has shown that a superconducting phase  exists in those samples prepared with Se deficiency (or Fe excess)\cite{hsu,wu4,deguchi}, although, it is established that the $T_c$ increases for samples close to stoichiometry \cite{williams}. On the other hand, core-level spectroscopic as well as theoretical studies on the electronic structure of $\textrm{FeSe}_{\delta}$ for various values of $\delta$ reveal that the total density of states (DOS) is mainly dominated by the Fe $3d$ states that hybridize with Se $4p$ states close to the Fermi level, and when Fe interstitials are introduced, the density of $3d$ states is considerably enhanced \cite{kurmaev}. From the perspective of the crystalline structure, this behavior is attributed to a lattice distortion as $\delta$ is varied \cite{lee2, chen7}. These results indicate that, in these materials, $3d$ charge carriers are itinerant in character and mostly responsible for the superconducting properties. Despite these great efforts, exactly how the Se deficiency affects the charge carriers remains uncertain, and hence, more experimental investigations on the electronic structure are needed.

One route that can shed light on this issue is to study whether chemical substitutions or doping, either to the Se-site or the Fe-site, have any effect on enhancing or suppressing superconductivity. Previous reports \cite{deguchi,fang} on the superconducting properties of the system $\textrm{FeSe}_{1-x}\textrm{Te}_x$ showed that a $T_c$ appears even for values of $x$ as large as 0.9. Hall studies demonstrated that the charge carriers in $\textrm{FeTe}_{0.82}$ are mainly electrons and that the structure transition, when replacing Se by Te, may lead to a change in the electronic band structure and/or the variation in the scattering rate of charge carriers. Although the crystalline and superconducting properties of $\textrm{FeSe}_{1-x}\textrm{Te}_x$ have been widely explored, \cite{wu4,yeh,yeh1,gresty,grechnev,viennois1,margadonna,mcqueen} an exhaustive examination of its electronic properties is still lacking. Of great importance is to settle the question of whether or not the members of the 11 family are strongly correlated materials. This topic has been widely discussed in the literature for the case of iron-pnictides, and a large sector of researchers seems to converge to the view that these materials behave as weakly (or at most moderately) correlated systems \cite{kurmaev1, kurmaev2, mcleod5, singh,singh1,aichhorn1,yang1}. Nonetheless, the issue for the 11 compounds is still quite ambiguous. Previous theoretical and X-ray spectroscopic studies \cite{kurmaev,subedi} realized on samples of $\textrm{FeSe}_{\delta}$ have shown that these compounds behave as weakly correlated systems, sharing great similarities with the iron-pnictides. By contrast, some researchers have reported experimental \cite{tamai, pourret,homes} evidence of strong electron correlations in samples with Te substitution. Aichhorn et al. performed theoretical studies on the system FeSe including screened Coulumb interactions within the context of the dynamical mean-field approach \cite{aichhorn}. They demonstrated the appearance of lower Hubbard bands associated with strong electron correlations. However, no experiments exist in the literature to corroborate this prediction. 

From our standpoint, detailed and systematic experimental research is crucial to elucidate the role played by chalcogen atoms in the electronic properties of the Fe-chalcogenide superconductors that will ultimately lead not only to a better understanding of their electronic correlation effects but also to a clearer picture on the origin of superconductivity. X-ray emission (XES) and absorption spectroscopy (XAS) techniques are excellent tools to investigate the electronic properties of these materials since they probe the occupied and unoccupied states, respectively. In this work we extend the scope of our previous research \cite{kurmaev} and evaluate the effect on the valence states of FeSe superconductor when Se is gradually replaced by Te (in another work we discuss the case of chemical substitution of iron by cobalt \cite{iperez14a}). Particularly, we study the electronic properties for five different stoichiometries in the system $\textrm{Fe}\textrm{Se}_{1-x}\textrm{Te}_x$ (with $x=0.00$, 0.25, 0.50, 0.75, and 1.00) using density functional theory (DFT) and X-ray spectroscopy. With these tools we extract valuable physics on the valence states, the spin state, and the strength of electronic correlations. The partial density of states reveals a shift of the $p$ states of FeTe when compared to the rest of the series that we think can be correlated to the absence of superconductivity in FeTe. 

\section{Experimental and Calculations}\label{expcal}
\subsection{Synthesis and Crystalline Structure of $\textrm{Fe}\textrm{Se}_{1-x}\textrm{Te}_x$}
\label{crystasyn}
Samples were synthesized via solid state reaction using high purity chemical Alfa Aesar powders of Fe 99.998\%, Se 99.5\%, and Te 99.99\%. The reagents were mixed at the appropriate quantities using a mortar and a ball mill. Then the mixture was introduced in quartz tubes and sealed in argon atmosphere at $10^{-2}$ Torr. The tubes were then introduced into a furnace at 700 $^\circ$C for 5 days; then the temperature was decreased to 420 $^\circ$C for 2 more days. After this time, the tubes were cooled down to room temperature. Finally, the samples were ground again and kept in sealed containers. The crystalline structure of the samples was determined by X-ray diffraction (XRD) with a Siemens D5000 X-ray diffractometer using a Cu K$\alpha$ source with a Ni filter. Operation parameters were set to 34 kV and 25 mA. Phase identification was realized using the ICSD 2004 database through the program Match (version 2.1f). Intensities were measured at room temperature in 0.02$^\circ$ steps, in the 6$^\circ$ - 130$^\circ$ 2$\theta$ range (12 h). The Rietveld refinement was carried out using the program MAUD v2.33. Further details on sample characterization for magnetization and transition temperature are described elsewhere.\cite{gomez} 
\begin{figure}[tp]
\begin{center}
\includegraphics[width=12cm]{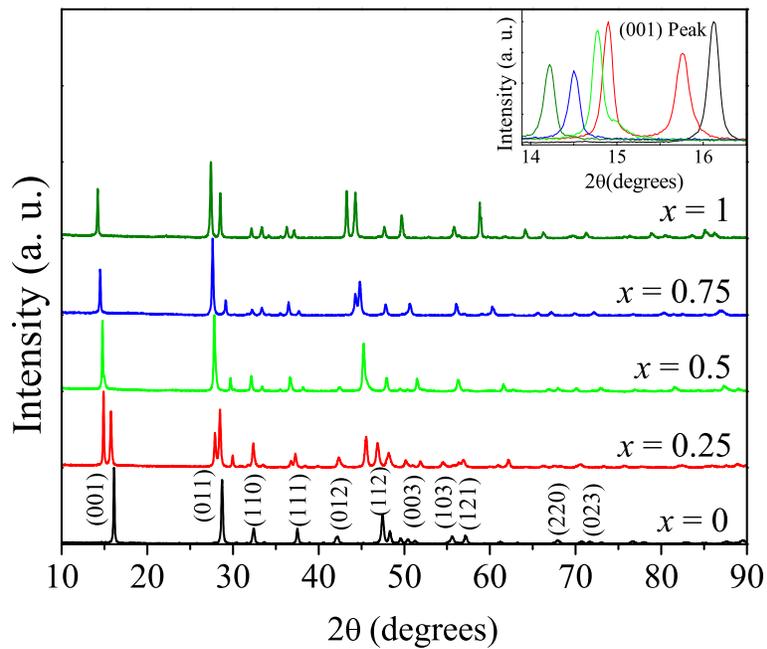}
\caption{X-ray diffraction spectra for the system $\textrm{Fe}\textrm{Se}_{1-x}\textrm{Te}_x$ with $x=0-1$. The patterns match the $P4/nmm$ space group. Inset shows a zoom of the (001) peak.}
\label{fetesexrd}
\end{center}
\end{figure}

\begin{figure}[hp]
\begin{center}
\includegraphics[width=11cm]{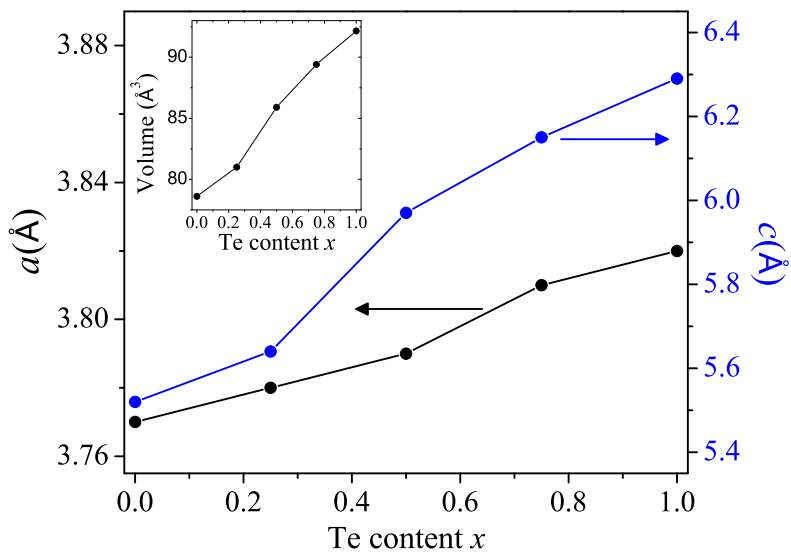}
\caption{Lattice parameters $a$ (black) and $c$ (blue). Inset shows the lattice volume.}
\label{latticepar1}
\end{center}
\end{figure}
 \begin{table}[tp]
   \caption{Crystallographic information of $\textrm{Fe}\textrm{Se}_{1-x}\textrm{Te}_x$.}
  \centering 
  \begin{tabular}{c|ccccc}
  \hline
$x$ & 0&0.25 &    0.5 &  0.75 &  1
  \\ \hline
  \footnotesize{\%$\textrm{Fe}\textrm{Se}_{1-x}\textrm{Te}_{x}$}  ($P4/nmm$)&   \footnotesize{82.8} & \footnotesize{62.8} & \footnotesize{91.6} & \footnotesize{93.10}  &  \footnotesize{94.9} \\ 
\footnotesize{\%$\textrm{Fe}\textrm{Se}$}  ($P6_3/mmc$)&   \footnotesize{18.21} & \footnotesize{1.8} & \footnotesize{8.4} & \footnotesize{-}  &  \footnotesize{-} \\ 
\footnotesize{\%$\textrm{Fe}\textrm{Te}$} ($P4/nmm$) &   \footnotesize{-} & \footnotesize{35.4} & \footnotesize{-} & \footnotesize{-}  &  \footnotesize{-} \\ 
\footnotesize{\%$\textrm{Fe}\textrm{Te}_2$}  ($Pnnm$)&   \footnotesize{-} & \footnotesize{-} & \footnotesize{-} & \footnotesize{6.9}  &  \footnotesize{5.1} \\  \hline
  
    \footnotesize{Bond length (\AA)} &   \footnotesize{} & \footnotesize{} & \footnotesize{} & \footnotesize{}  &  \footnotesize{} \\ 
  \footnotesize{Fe-Fe:} &   \footnotesize{2.6680} & \footnotesize{2.6781} & \footnotesize{2.6811} & \footnotesize{2.6960}  &  \footnotesize{2.7061} \\ 
  \footnotesize{(Se/Te)-Fe:} &  \footnotesize{2.3712} & \footnotesize{2.3970} & \footnotesize{2.4510} & \footnotesize{2.4881}  &  \footnotesize{2.5592} \\   \hline
  
    \footnotesize{Bond angle (deg)} &  \footnotesize{} & \footnotesize{} & \footnotesize{} & \footnotesize{}  &  \footnotesize{} \\ 
  \footnotesize{(Se/Te)-Fe-(Se/Te)} &   \footnotesize{105.43} & \footnotesize{112.11} & \footnotesize{113.70} & \footnotesize{114.41}  &  \footnotesize{115.01} \\ 
  \footnotesize{Fe-(Se/Te)-Fe} &  \footnotesize{68.47} & \footnotesize{67.91} & \footnotesize{66.32} & \footnotesize{65.60}  &  \footnotesize{65.01} \\  \hline
  
    \footnotesize{B(\AA)$^2$} &   \footnotesize{} & \footnotesize{} & \footnotesize{} & \footnotesize{}  &  \footnotesize{} \\ 
  \footnotesize{Fe} &   \footnotesize{0.78} & \footnotesize{0.37} & \footnotesize{0.42} & \footnotesize{0.49}  &  \footnotesize{0.49} \\ 
    \footnotesize{Se/Te} &   \footnotesize{1.77} & \footnotesize{0.67} & \footnotesize{0.82} & \footnotesize{0.49}  &  \footnotesize{0.49} \\   \hline
  
    \footnotesize{Occupation factor N} &   \footnotesize{} & \footnotesize{} & \footnotesize{} & \footnotesize{}  &  \footnotesize{} \\ 
  \footnotesize{Fe} &   \footnotesize{0.98} & \footnotesize{0.97} & \footnotesize{0.98} & \footnotesize{0.96}  &  \footnotesize{0.96} \\ 
  \footnotesize{Se} &   \footnotesize{1.01} & \footnotesize{0.84} & \footnotesize{0.5} & \footnotesize{0.26}  &  \footnotesize{0.00} \\ 
    \footnotesize{Te} &   \footnotesize{0.00} & \footnotesize{0.16} & \footnotesize{0.5} & \footnotesize{0.77}  &  \footnotesize{1.03} \\  \hline
 
   \hline
\end{tabular}
 \label{fesetetable}
\end{table}

In \ref{fetesexrd} we display the XRD spectra for our system and in \ref{fesetetable} we show the results of the Rietveld refinement. The crystal analysis reveals that all stoichiometries exhibit the tetragonal phase with space group $P4/nmm$ corresponding to $\textrm{FeSe}_{1-x}\textrm{Te}_x$. The diffraction pattern for $x= 0.25$ shows peak splitting, implying the coexistence of two phases with tetragonal structure: the phase FeTe and the phase $\textrm{FeSe}_{1-x}\textrm{Te}_x$, consistent with previous reports \cite{fang,yeh,mizuguchi}. This suggests that the structure of FeSe is fundamentally different from the structure of FeTe despite the observation that both can be indexed by the same tetragonal lattice. The fact that the second phase has a $T_c$ of about 12 K implies that its composition is different from the nominal composition. From the Rietveld refinement (refer to \ref{fesetetable}), 62.8\% of the ternary phase is formed giving a composition $\textrm{FeSe}_{0.843}\textrm{Te}_{0.157}$. There is also a clear non-linear increase in the lattice parameters in proportion to Te concentration due to the larger ionic radius of Te (see \ref{latticepar1}). The expansion of the lattice is asymmetrical, with the expansion rate along the $c$-direction much faster. The change in the slope delimits the crossover between the two phases that occurs at $x=0.25$. The variation of the parameters is also reflected in the (001) peak shift to lower $2\theta$ values (see inset in \ref{fetesexrd}). The samples with $x$ = 0.75 and 1.00 additionally display the orthorhombic phase corresponding to $\textrm{FeTe}_2$ as an impurity in the system (spatial group $Pnnm$). The appearance of this phase seems to affect the superconducting state since superconductivity is destroyed for Te concentrations larger than 90\%.\cite{fang,yeh,gomez} It is worth mentioning that the samples with $x= 0.00$, 0.25 and 0.50 exhibit impurities corresponding to the NiAs-type hexagonal FeSe (spatial group $P6_3/mmc$). Apparently, this phase plays no significant role in the superconducting properties of FeSe; however, we discovered\cite{iperez14a} that in the $\textrm{Fe}_{1-y}\textrm{Co}_y\textrm{Se}$ system the introduction of Co into the host lattice favors this phase and dominates the system for $y>0.38$. In this case, the $T_c$ diminishes rapidly as $y$ increases\cite{mizuguchi}, going from 10 K for $y=0$ to 5 K for $y=0.1$, and for $y>0.15$ the superconducting state is completely destroyed. In contrast, we observe (see \ref{fesetetable}) that the effect of substituting Se by Te is to eliminate the hexagonal phase and enhance the tetragonal one. Previous crystallographic analysis \cite{fang,yeh} showed that the angle $\gamma$ as a function of Te concentration varies in the same proportion as the $T_c$. This strongly suggests that the $T_c$ can be correlated to the lattice distortions more than the Fe-Fe distance in the Fe plane which increases monotonically as $x$ increases. By comparing the superconducting and crystal properties of $\textrm{Fe}_{1-y}\textrm{Co}_y\textrm{Se}$ with those of $\textrm{FeSe}_{1-x}\textrm{Te}_x$, we realize that the tetragonal structure is determinant in stabilizing FeSe in the superconducting state. Thus, while Fe substitution by Co enhances the hexagonal phase and ultimately destroys superconductivity for $y>0.15$, the introduction of Te in FeSe eliminates the hexagonal phase but increases the lattice size and distorts the tetragonal lattice to such an extent that the orthorhombic phase starts to emerge for $x>0.9$ and finally, the superconducting state in FeTe is again suppressed. Indeed, in the following sections we shall discuss how the electronic properties of FeSe are influenced by the introduction of Te. Our DFT calculations indicate structural fluctuations between tetragonal and orthorhombic phases and reveal a narrowing of the valence band for $x=1$. The density of states shows variations in the $p$ states of FeTe that can be correlated to the appearance of the orthorhombic phase and the suppression of superconductivity.
 
\subsection{X-ray Spectroscopy Measurements}
 In this research we measured the Fe $L_{2,3}$ non-resonant XES, resonant inelastic X-ray scattering (RIXS), and XAS spectra for the five stoichiometries described in the previous section. The measurements were carried out at the soft X-ray fluorescence endstation at Beamline 8.0.1 of the Advanced Light Source (ALS) at Lawrence Berkeley National Laboratory. The endstation has a Rowland circle geometry X-ray spectrometer with spherical gratings and an area-sensitive multichannel detector \cite{jia}. The instrumental resolving power (E/$\Delta E$) for XES spectra was approximately $10^3$. For the XAS measurements we used the surface-sensitive total electron yield (TEY) mode. The instrumental resolving power for all XAS measurements was about $5\times10^3$. During the measurement sessions the samples were placed under high vacuum ($10^{-8}$ Torr) and measured at room temperature. Emission spectra were normalized with respect to the Fe $L_3$ peak. Absorption spectra were divided by the incident photon current and then normalized by the Fe $L_3$ peak as well. The excitation energies for the RIXS measurements were determined from the Fe $2p$ XAS measurements. They corresponded to the location of the $L_2$  and $L_3$ peaks, an energy between them, and an energy well above the $L_2$ threshold for the non-resonant XES.

\subsection{Calculation Details}
Density functional theory with the full-potential linearized augmented plane-wave (LAPW) method as
implemented in the WIEN2k code \cite{blaha} was used to compute the electronic structure of our systems. For the exchange correlation potential we employed the generalized gradient approximation in the Perdew-Burke-Ernzerhof variant \cite{perdew}. We generated a $12 \times12\times7$ k-mesh to perform the Brillouin zone integrations, and for the expansion of the basis set we chose $R^{min}_{MT} K_{max}=7$ (the product of the smallest of the atomic sphere radii $R_{MT}$ and the plane wave cutoff parameter $K_{max}$). The radii of the muffin-tin spheres for the atoms were chosen so that the neighboring spheres were nearly touching. The values used were: $R_{\textrm{Fe}}=2.21$, $R_{\textrm{Se}}=1.96$ and $R_{\textrm{Te}}=2.00$. For the calculations we  used the experimental values of the lattice parameters as determined from XRD. To simulate Te substitution for the stoichiometries with $x=$0.25, 0.50, and 0.75, we generated $2\times 2\times 1$ supercells. In these supercells, a substitution of two Se atoms by two Te atoms represents a 25\% substitution. The space group for these structures were, respectively: orthorhombic $Pmm2$, tetragonal $P4mm$, and again $Pmm2$. The supercell calculations were not optimized because the experimental details are well known \cite{viennois1,margadonna,mcqueen}. For FeSe and FeTe we considered tetragonal structures belonging to the $P4/nmm$ space group. In all cases energy convergence of 0.0001 Ryd, charge convergence of 0.001 $e$ and cutoff energy between core and valence states of -6 Ryd were chosen.

XES spectra were also calculated using the XSPEC package implemented in WIEN2k \cite{schwarz}. The package calculates the spectra based on the dipole allowed transitions which are then multiplied with a radial transition probability and the partial densities of states (PDOS).

\section{Analysis and Discussion}\label{analdis}
\subsection{Density of States}\label{dosi}
In a previous publication we studied the electronic properties of FeSe with Fe excess and Se deficiency using DFT and X-ray spectroscopy \cite{kurmaev}. The experimental evidence showed that FeSe is at most a moderate correlated material and that the bulk of 3$d$ states is concentrated around the Fermi energy ($E_F$). We also showed that non-stoichiometric FeSe raises the number of states at the Fermi level N($E_F$). The band structure for stoichiometric FeSe and FeTe was also studied using DFT some time ago by Subedi and Ma et al. \cite{ma,subedi}. Their investigations showed that the DOS and Fermi surface of these materials share some similitudes with Fe-As superconductors. In the present study we performed similar calculations, but we now include the experimental structures for four novel samples with several Te concentrations and support our results with X-ray spectroscopy. To the best of our knowledge there is no other work in the literature studying the DOS for intermediate values of $x$. This approach not only will prove to be useful in monitoring the evolution of the DOS as a function of Te substitution but also will serve to detect any trends that could be correlated with the superconducting properties.
\begin{figure}[tp]
\begin{center}
\includegraphics[width=10cm]{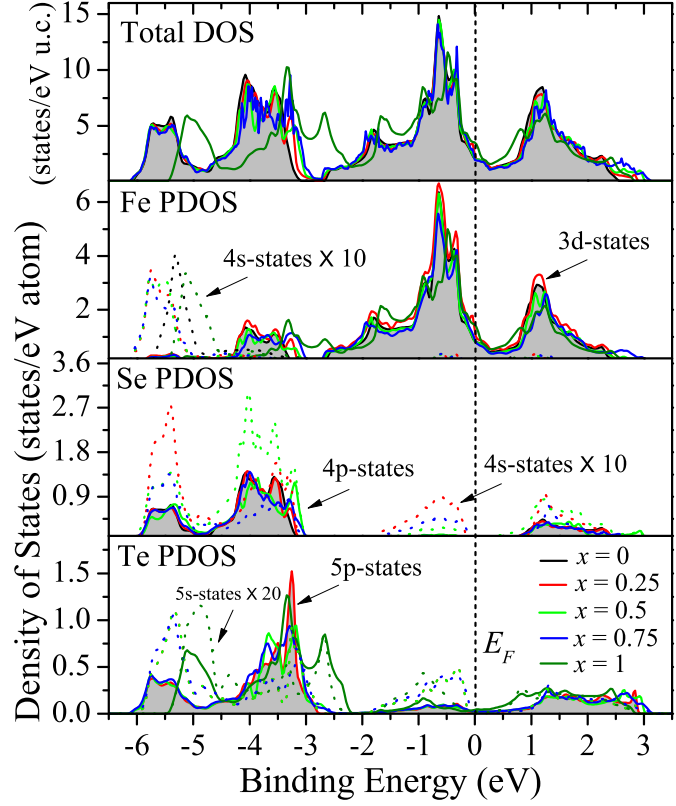}
\caption{Density of states for $\textrm{FeSe}_{1-x}\textrm{Te}_x$. The Fermi energy ($E_F$) is set at zero energy and is marked by a vertical dotted line. For visualization the PDOS for Fe and Se $s$-states are multiplied by a factor of 10 and Te $s$-states by a factor of 20 (dotted spectra).}
\label{DOSFTotFeSeTe1}
\end{center}
\end{figure}

In \ref{DOSFTotFeSeTe1} we display the DOS for $\textrm{FeSe}_{1-x}\textrm{Te}_x$. Near the Fermi level the total DOS for stoichiometries with $x<1$ is mostly dominated by a band of $3d$ Fe electrons extending from 0.0 eV to $-2.7$ eV. After a small gap of approximately 0.3 eV, a second band of hybridized $3d$ Fe and $4p$ Se / $5p$ Te electrons extends from $-3.0$ eV to $-6.1$ eV. The Fe $3d$ and the chalcogen $p$ states are quite similar for these compositions, the main difference comes from slight variations in the $4p$ / $5p$ states between -3 eV and -4 eV. Moreover, the Fe $4s$ and the chalcogen $s$ states show evident differences, although no clear trend is observed and their contribution in this energy range is insignificant. 
    \begin{table}[bp]
   \caption{Number of states at the Fermi level of $\textrm{FeSe}_{1-x}\textrm{Te}_x$.}
  \centering 
  \begin{tabular}{c|c|c}
  \hline
Stoichiometry & Space  & N($E_F$) \\
$x$&group& \footnotesize{(states/eV f.u.)}
  \\ \hline
\footnotesize{0.00} & \footnotesize{$P4/nmm$} & \footnotesize{1.04}  \\
\footnotesize{0.25} &   \footnotesize{ $Pmm2$} & \footnotesize{1.22}   \\
\footnotesize{0.50} & \footnotesize{$P4mm$ } & \footnotesize{1.06}  \\
\footnotesize{0.75} & \footnotesize{$Pmm2$} & \footnotesize{0.99}    \\
\footnotesize{1.00} & \footnotesize{$P4/nmm$} & \footnotesize{1.75} \\
 \hline
\end{tabular}
 \label{values}
\end{table}
 Turning now our attention to the total DOS for the system FeTe, we see that the valence band extends from $-5.4$ eV to 2.8 eV, slightly narrower than the other stoichiometries. This band has contributions from Fe $3d$ and Te $5p$ states and is also dominated by the $3d$ states near the Fermi level. The $3d$ states show no gap if compared to the other stoichiometries, and the $p$ states at the bottom of the valence band emerge about 0.5 eV to the right. This is important if we keep in mind that superconductivity is suppressed for FeTe and the orthorhombic phase appears as an impurity. Another distinctive feature is the N($E_F$); the corresponding values along with the crystalline structures are given for reference in \ref{values}. FeTe has the highest value but still lower than the value of 2.62 states per eV per formula unit (f.u.) obtained in $\textrm{LaFeAs}(\textrm{O}_{1-x}\textrm{F}_x)$ \cite{singh}. Surprisingly, in our previous work we reported that Fe excess or Se deficiency at least doubles the number of 3$d$ states at $E_F$ (from 0.87 states/eV/atom to 2.19 states/eV/atom) which is comparable to the value reported in Fe-pnictide materials \cite{singh,singh1}. Notice that, for stoichiometries with $x<1$, N($E_F$) remains close to that of FeSe. An interesting finding we spot is that the overall shape of the total DOS is barely unaffected by Te substitution up to $x=0.75$. Since superconductivity shows up for $x<0.9$, our results suggest that the distribution of $p$ states might be fundamental for superconductivity in this system. The shift of the $5p$ states of FeTe along with the elimination of the energy gap cannot be caused by charge-carrier doping since Se$^{2-}$ and Te$^{2-}$ have the same valence; instead, the effect can be explained by both the lattice distortion induced by the larger ionic radius of Te$^{2-}$ and the appearance of the orthorhombic phase (refer back to \ref{crystasyn}). This is plausible since the calculations also predict the existence of the orthorhombic phase for $x=0.25$ and 0.75. In general, we observe that the DOS for $\textrm{FeSe}_{1-x}\textrm{Te}_x$ displays some generic characteristics that resemble the DOS of most Fe-pnictide superconductors \cite{kurmaev1,mcleod5}; namely, a major contribution comes from the Fe $3d$ states in the neighbourhood of the Fermi level and a modest contribution from the chalcogen $s$ and $p$ states that decreases noticeably with energy near the $E_F$. Because it is well known that the Fe-pnictides are weakly correlated materials, this similarity suggests that $\textrm{FeSe}_{1-x}\textrm{Te}_x$ can be also considered at most as a moderately correlated system. In fact, in the following subsections we discuss this topic at length and give some experimental evidence to back up this view.

\subsection{RIXS Measurements and the Spin State}

RIXS is a powerful technique that can probe the valence states of transition metal atoms and provide valuable information about the electronic structure of a system. One of the most important problems about the FeSe system is the determination of its spin state and magnetic order. In this section we will use RIXS to extract information on the electronic structure and estimate the spin state of our system that can be correlated to the magnetic order. 

\begin{figure}[tp]
\begin{center}
\includegraphics[width=9cm]{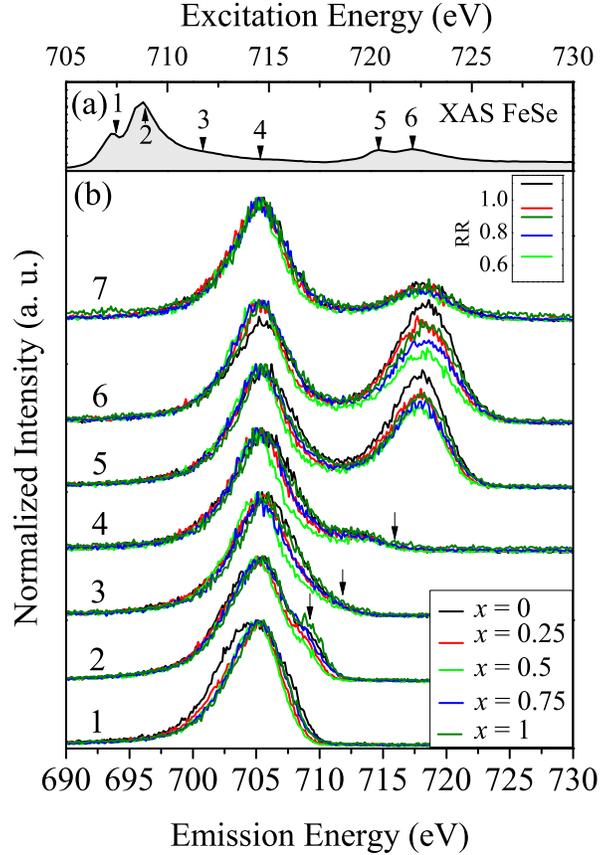}
\caption{Fe $L_{2,3}$ absorption spectrum (a) for FeSe and RIXS spectra (b) for $\textrm{FeSe}_{1-x}\textrm{Te}_x$. The excitation energies used to collect the RIXS spectra are indicated by arrows in the XAS spectrum. Arrows in spectra 2-4 indicate the emission by elastic scattering tracking the excitation energy. Inset shows the values for the resonant ratio (RR) $I(L_2)/I(L_3)$ for spectra 6.}
\label{rixsfese}
\end{center}
\end{figure}

In \ref{rixsfese}(a) we show the Fe $L_{2,3}$ absorption structure of FeSe, whereas the RIXS spectra for $\textrm{FeSe}_{1-x}\textrm{Te}_x$ are shown in \ref{rixsfese}(b). The excitation energies used to collect the RIXS spectra are marked with arrows in the absorption spectra (nos. 1-6). The non-resonant XES measurements were taken at an excitation energy of 740 eV far beyond the absorption threshold (spectra no. 7). 
These curves present the two main $L_{2,3}$ fluorescence bands generated by the spin-orbital splitting. The peaks are located around 705 eV and 718 eV, which correspond to $L_3$ and $L_2$ emission lines produced from the transitions $3d4s\to 2p_{3/2}$ and $3d4s\to 2p_{1/2}$, respectively. As seen in the RIXS spectra 2-4, small shoulders (marked with arrows) start to appear at energies of approximately 710 eV and 715 eV. These energies track the excitation energies and are caused by elastic scattering. The RIXS and the non-resonant XES spectra for stoichiometries with $x>0$ are basically featureless since no signs of constant energy loss features, which are usually associated with charge transfer or $d$-$d$ excitations, are detected. The $L_3$ peak resembles that of pure Fe and the absence of satellites indicates that $3d$ electrons are mainly itinerant in character \cite{schwarz,kurmaev3,gao}. These findings are quite similar to those reported for Fe metal and Fe-pnictide superconductors \cite{kurmaev1,kurmaev2,mcleod5,yang1} and considerably contrast with those found on most transition-metal oxides whose spectral structure is very rich and reflects strong correlation effects \cite{prince}. These observations indicate that the non-resonant $L_{2,3}$ XES probes the partial DOS and that $\textrm{FeSe}_{1-x}\textrm{Te}_x$ behaves as weakly correlated system. 

A closer inspection of the RIXS spectra reveals a subtle peculiarity that can be used to obtain important information on the spin state. Its resonant excitation at the $L_2$ threshold (spectra 5 and 6) makes the $L_2$ peak quite a bit higher in comparison to the $L_2$ peak in the other compounds with the same excitation energy. In other words, the ratio of the intensities of the $L_2$ to $L_3$ peak, for these excitation energies, is larger than the same ratio in the other stoichiometries. We calculated the $I(L_2)/I(L_3)$ ratio for the RIXS spectra number 6 --- we shall call it hereafter the resonant ratio ($RR$). The corresponding values are shown in the upper inset in \ref{rixsfese} (b). We see that FeSe has a much higher $RR$ ($=1.1$) than the rest of the stoichiometries. Prince et al. observed similar variations in the $RR$ of FeO and of FeS$_2$ \cite{prince}. In both cases the differences in the $RR$s were ascribed to differences in the intrinsic Coster-Kronig (C-K) rate of each sample, with a higher rate for FeS$_2$. In turn, the differences in the C-K probability were related to the spin state of the systems (details which can be consulted in the corresponding paper). Accordingly, the authors formulated a rule for resonant emission very similar to that proposed for absorption \cite{thole} and for non-resonant emission \cite{yablonskikh}. The rule states that the $RR$ is higher for high spin ground states than for low spin ground states. Thus the resonant spectra can be used to characterize the magnetic state of a material. According to this rule, magnetic $3d$ materials tend toward high $RR$s whereas nonmagnetic materials tend to low ones. The authors reported that FeO is antiferromagnetic with Fe in the Fe$^{2+}$ high spin state (S=2 in cubic symmetry), with $RR=1.35$, while FeS$_2$ is nonmagnetic with Fe in the Fe$^{2+}$ low spin state (S=0 in octahedral symmetry) with $RR= 0.47$. Therefore, taking as reference these systems and applying this rule to our system, Fe in FeSe should be in a higher spin state than the spin state for the rest of the series. There is a controversy about the spin state of Fe atoms in FeSe and FeTe. The theoretical considerations carried out by Ma et al. \cite{ma} presumed that FeTe is in a bicollinear antiferromagnetic order while FeSe is in a collinear antiferromagnetic one. Recent calculations \cite{wu3} on the magnetic ordering of FeSe using hybrid-exchange DFT showed that nonmagnetic order has higher ground state energy than the magnetic one, although temperature and pressure variations as well as lattice distortions (as those taking place in $\textrm{FeSe}_{1-x}\textrm{Te}_x$) could drive the material into different magnetic regimes. Using polarized Raman-scattering techniques, very recent experiments \cite{gnezdilov,gnezdilov1} showed that FeSe and FeTe undergo spin fluctuations as a function of temperature; manifesting higher spin states at high temperatures and driving FeSe (FeTe) from ferromagnetic (antiferromagnetic) to paramagnetic as the temperature increases. The estimations of the spin state at room temperature revealed that FeSe has a higher spin state (S=2) than FeTe (S=1). These findings are in qualitative agreement with both the spin-state rule and the values of the $RR$ for FeSe and FeTe found here. The values of the $RR$ for $x=0.25$ and 0.75 are closer to the value of FeTe, suggesting that these two stoichiometries have the same spin state as FeTe. The $RR$ for $x=0.5$ is much closer to the $RR$ of FeS$_2$ and consequently, a low spin state (S=0) can be assigned. 

\subsection{Non-resonant XES Analysis and Strength of Electronic Correlations}
Another important question that is still on the table is whether FeSe is a strongly correlated system from the point of view of the on-site Hubbard $U$. In this section we attempt to shed light on this issue.
\begin{figure}[tp]
\begin{center}
\includegraphics[width=10cm]{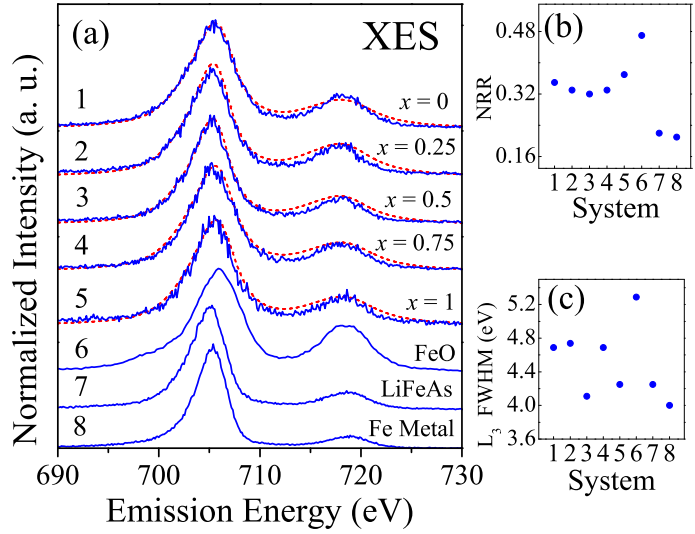}
\caption{(a) Non-resonant Fe $L_{2,3}$ XES spectra for $\textrm{FeSe}_{1-x}\textrm{Te}_x$, FeO, LiFeAs and Fe metal. (b) Non-resonant ratio (NRR) of the $L_{2,3}$ intensities. The intensity ratios were derived by computing the integral under the $L_{2,3}$ bands. (c) The full width at half maximum (FWHM) for the $L_3$ peaks from spectra in part (a). Solid spectra in blue represent measurements and dashed spectra in red are the calculations.}
\label{feseteXES}
\end{center}
\end{figure}
The metallicity of our system can be assessed by analyzing the non-resonant XES spectra since they unveil information about the occupied density of states around the Fermi level. To this end, in \ref{feseteXES}(a) we compare the simulated XES spectra with the measurements. The calculated spectra were Lorentzian- and Gaussian-broadened to account for lifetime and instrumental broadening, respectively, and were manually shifted to match the measured spectra. From the figure we can see that calculations are in great agreement with the measurements, although, due to broadening, no fine details are expected to be revealed. Despite this, the absence of secondary features not only suggests that the Fe $3d$ states do not form strong bonds with chalcogen states outside the $3d$ band but also that strong correlation effects such as on-site Hubbard $U$ correlations barely occur in this system. If this were the case, lower Hubbard bands in the DOS would manifest \cite{aichhorn} and the structure of the emission spectra would display additional features. To contemplate the possibility of a significant Coulumb repulsion, we also performed LDA+U calculations (not shown) for FeSe within the context of the self-interaction correction (SIC) and the Hubbard mean field (HMF) variants. However, as expected, the DOS was drastically influenced by the magnitude of the Coulumb parameter $U$, specifically for values greater than 0.5 eV. As $U$ was increased beyond 7 eV, the $3d$ states shifted to higher energies and prominent features on the non-resonant XES spectrum showed up (such as relatively large bandwidth broadening and shoulders around the $L_3$ peak), in clear conflict with our measurements. The fact that $U$ must be close to zero to reproduce with good degree of accuracy the emission spectra does not suggest that $\textrm{FeSe}_{1-x}\textrm{Te}_x$ is a strongly correlated system. Medici et al. \cite{medici} proposed that moderate to strong correlations in Fe-based superconductors can be driven by Hund's rule coupling $J$ rather than on-site Hubbard repulsion. This appears to be plausible since the $I(L_2)/I(L_3)$ ratios of the non-resonant emission of $\textrm{FeSe}_{1-x}\textrm{Te}_x$ suggest at most moderate correlations. The alternative was considered in the Fe-pnictides \cite{yang1,haule,johannes} albeit it was shown that relatively large values of the coupling parameter $J$ ($>$1 eV with $U$=0.9 eV) lead to the appearance of a high-energy shoulder on the $L_3$ absorption peak accompanied by a two-peak splitting, features which are absent in the absorption spectra of both the Fe-pnictides and our system (refer back to \ref{rixsfese}(a)). On the basis of these considerations, a naive estimation for the parameters $U$ and $J$ sets an upper limit of 0.5 eV and 1 eV, respectively.

In order to further evaluate the degree of correlation strength in our system, in \ref{feseteXES}(a) we compared the non-resonant XES spectra of $\textrm{FeSe}_{1-x}\textrm{Te}_x$ (nos. 1-5), FeO (no. 6),  LiFeAs (no. 7) and Fe metal (no. 8). First, we see that the height of the $L_2$ peak for Fe metal is smaller than the height of the same peak in the other materials. On the opposite extreme, the highest $L_2$ peak corresponds to FeO. For quantitative analysis we have calculated the ratios of the integrated intensities of the $L_2$ to $L_3$ emission, $I(L_2)/I(L_3)$, that is, the non-resonant ratio (NRR). This quantity reflects the statistical population of $2p_{1/2}$ and $2p_{3/2}$ energy levels, respectively. According to the one-electron picture, the NRR should be 1/2; however, in metals, C-K transitions considerably reduce the emission yield of the $L_2$ band. Consequently, the NRR can provide a qualitative probe of the metallicity and correlation strength of a system \cite{kurmaev3}. We note however that self-absorption effects are present and can neither be eliminated nor corrected\cite{kawai1,kawai2} and consequently, they may have a significant impact on the magnitude of both the NRR and the RR. However, a simple reflection shows that this is not the case. The XAS maximum occurs slightly above 708 eV (refer back to \ref{rixsfese}(a)) and the $L_3$ emission maximum at about 705 eV (see \ref{feseteXES}(a)). This means that on the right side of the emission band the emitted photons are experiencing increasing absorption with increasing emission energy when escaping the sample. The same holds true, although to an even weaker extent, for the $L_2$ band where the contrast is weaker. Ultimately self-absorption will result in a slightly distorted line shape on the high-energy side of the peaks. While we agree that the self-absorption effect is always present, it is also clear that it does not affect our preceding and forthcoming analysis since only the high-energy side of the profile is slightly suppressed. Furthermore, since the Fe concentration remains constant in all our measurements, the same small distortion will apply to all spectra, and therefore, the effect can be neglected.

From \ref{feseteXES}(b) we see that the NRRs of $\textrm{FeSe}_{1-x}\textrm{Te}_x$ are between Fe metal and the strongly correlated FeO, indicating that the $3d$ electrons in these materials exhibit somewhat localized character when compared to Fe metal (or LiFeAs). FeSe and FeTe are much closer to FeO suggesting that the $3d$ electrons are slightly more localized than in the others. These findings become relevant when they are confronted with those found in iron-pnictide superconductors where $3d$ electrons are mainly itinerant \cite{mcleod5,yang1}. The itinerant character in Fe-pnictide superconductors is exemplified by the LiFeAs system\cite{kurmaev2} since the value of its NRR is comparable to that of Fe metal. Not surprisingly, the XES spectrum of FeO is slightly shifted to higher energies and shows a typical satellite at the low-energy side of the $L_3$ band. These features are characteristic of most transition-metal oxides whose valence electrons are highly localized \cite{prince,laan}. This analysis supports the view that $\textrm{FeSe}_{1-x}\textrm{Te}_x$ can be considered at most as a moderately correlated system.

To conclude our discussion, we observe that the width of the $L_3$ band is wider than that in the other systems. A good quantification of the bandwidth could be given by calculating the full width at half maximum (FWHM). The FWHM of $L_3$ (shown in \ref{feseteXES}(c)) of $\textrm{FeSe}_{1-x}\textrm{Te}_x$ is closer to that of Fe metal than to that of FeO. This indicates that most of the $3d$ Fe electrons of $\textrm{FeSe}_{1-x}\textrm{Te}_x$ are to some extent concentrated in a narrow band close to the Fermi level, just as the calculations of the previous section predicted. For metallic compounds, it was found that the Fe $3d$ bandwidth should decrease with increasing Fe-Fe distance \cite{kurmaev2}. For $\textrm{FeSe}_{1-x}\textrm{Te}_x$, however, we do not identify any clear trend in the bandwidth that could be correlated to the variation of Fe-Fe distance as the level of Te varies (see \ref{fesetetable}). 

\section{Conclusions}\label{concl}
We have investigated the crystalline and electronic properties of $\textrm{FeSe}_{1-x}\textrm{Te}_x$ (with $x=0-1$) using XRD, X-ray spectroscopy, and DFT calculations. The analysis performed on the crystal structure of the $\textrm{FeSe}_{1-x}\textrm{Te}_x$ system indicates that the tetragonal structure plays an important role in favoring the superconducting state in FeSe. We discovered that Te eliminates the hexagonal phase but increases the lattice size due to the larger ionic radius of Te that finally destroys superconductivity for $x=1$ where the orthorhombic phase shows up. This fact can be correlated to the variations of the $p$ states of FeTe spotted in the projected DOS. The calculations for the band structure showed that the Fe $3d$ states dominate the DOS in the vicinity of the Fermi energy. A comparison of our results with previous studies on the band structure of Fe-pnictides tells us that the main features found here are common to most Fe-based superconductors; namely, Fe $3d$ states dominate near the Fermi level and the $3d$ and the $4p$ / $5 p$ states hybridize at the bottom of the valence band. 

On the other hand, using the RIXS technique, we also estimated the spin state of $\textrm{FeSe}_{1-x}\textrm{Te}_x$. We have found that the spin state fluctuates as a function of $x$ with FeSe in the highest spin state (S=2), in agreement with polarized Raman scattering measurements. 

Finally, we assessed the degree of electronic correlations. On one hand, the Fe $L_3$ XES simulations were in good agreement with the experimental counterpart supporting the view of weak electronic correlations. On the other hand, the NNR suggested that $\textrm{FeSe}_{1-x}\textrm{Te}_x$ behaves, at most as a moderately correlated system, with FeSe and FeTe the most correlated materials. Overall, we found that our system is more correlated than the iron-pnictide superconductors.

\begin{acknowledgement}

The authors gratefully acknowledge support from the Natural Sciences and Engineering Research Council of Canada (NSERC) and the Canada Research Chair program. This work was done with partial support from CONACYT Mexico under Grant 186142 and from Programa de Apoyo a Proyectos de Investigaci\'on e Innovaci\'on Tecnol\'ogica (PAPITT), UNAM under Project IN115410. The Advanced Light Source is supported by the Director, Office of Science, Office of Basic Energy Sciences of the U. S. Department of Energy, under Contract DE-AC02-05CH11231. The computational part of this research was enabled using computing resources provided by WestGrid and Compute/Calcul Canada. The authors are indebted to the reviewers for helpful comments and suggestions that undoubtedly improved the quality of this research.

The authors declare no competing financial interest.

\end{acknowledgement}

\bibliography{achemso-demo}

\begin{thebibliography}{99}

\bibitem {kamihara} Kamihara, Y.; Watanabe, T.; Hirano, M.; Hosono, H. Iron-Based Layered Superconductor $\mathrm{La}[\mathrm{O}_{1-x}\mathrm{F}_x]\mathrm{FeAs}$ (x=0.015-0.12) with $T_c=26$ K.
{\it J. Am. Chem. Soc.} {\bf 2008}, {\it 130}, 3296-3297

\bibitem {rotter} Rotter, M.; Tegel, M.; Johrendt, D.; Schellenberg, I.; Hermes, W.; P\"ottgen, R. Spin-Density-Wave Anomaly at 140 K in the Ternary Iron Arsenide $\textrm{BaFe}_2\textrm{As}_2$. {\it Phys. Rev.} B {\bf 2008}, {\it 78}, 020503-4

\bibitem {wang5} Wang, X. C.; Liu, Q. Q.; Lv, Y. X; Gao, W. B.; Yang, L. X.; Yu, R. C.; Li, F. Y.; Jin, C. Q. The Superconductivity at 18 K in LiFeAs System. {\it Solid State Commun.}  {\bf 2008}, {\it 148}, 538-540

\bibitem {hsu} Hsu, F. C.; Luo, J.-Y.; Yeh, K.-W.; Chen, T. K.; Huang, T.-W.; Wu, P. M.; Lee, Y.-C.; Huang, Y. L.; Chu, Y. Y.; Yan, D. C.; et al. Superconductivity in the PbO-type Structure $\alpha$-FeSe.
{\it Proc. Natl. Acad. Sci. U. S. A.}  {\bf 2008}, {\it 105}, 14262-14264

\bibitem{ma} Ma, F.; Ji, W.; Hu, J.; Lu, Z. Y.; Xiang, T. First-Principles Calculations of the Electronic Structure of Tetragonal $\alpha$-$\textrm{FeTe}$ and $\alpha$-$\textrm{FeSe}$ Crystals: Evidence for a Bicollinear Antiferromagnetic Order. 
{\it Phys. Rev. Lett.} {\bf 2009}, {\it 102}, 177003-5

\bibitem{wu4} Wu, M. K.; Yeh, K. W.; Hsu, H. C.;Huang, T. W.; Chen, T. K.; Luo, J. Y.;  Wang, M. J.; Chang, H. H.; Ke, C. T.; Mo, M. H.; et al. The Development of the Superconducting Tetragonal PbO-type FeSe and Related Compounds.  {\it Phys. Status Solidi} B {\bf 2010}, {\it 247}, 500-505

 \bibitem{deguchi} Deguchi, K.; Takano, Y.; Mizuguchi, Y. Physics and Chemistry of Layered Chalcogenide Superconductors.  {\it Sci. Technol. Adv. Mater.} {\bf 2012}, {\it 13}, 054303-11
 
\bibitem{williams} Williams, A. J.; McQueen, T. M.; Cava, R. J. The Stoichiometry of FeSe. {\it Solid State Commun.} {\bf 2009}, {\it 149}, 1507-1509

\bibitem {kurmaev} Kurmaev, E. Z.; McLeod, J. A.; Skorikov, N. A.; Finkelstein, L. D.; Moewes, A.; Korotin, M. A.; Izyumov, Y. A.; Xie, Y. L.; Wu, G.; Chen, X. H. Structural Models of $\textrm{FeSe}_x$. {\it J. Phys.: Condens. Matter} {\bf 2009}, {\it 21}, 435702-6

\bibitem {lee2} Lee, K. W.; Pardo, V.; Pickett, W. E. Magnetism Driven by Anion Cacancies in Superconducting $\alpha$-$\textrm{FeSe}_{1-x}$.  {\it Phys. Rev.} B {\bf 2008}, {\it 78}, 174502-5

\bibitem{chen7} Chen, C. L.; Rao, S. M.; Dong, C. L.; Chen, J. L.; Huang, T. W.; Mok, B. H.; Ling, M. C.; Wang, W. C.; Chang, C. L.; Chan, T. S. X-ray Absorption Spectroscopy Investigation of the Electronic Structure of Superconducting $\textrm{FeSe}_x$ Single Crystals. {\it Europhys. Lett.} {\bf 2011}, {\it 93}, 47003-5

\bibitem {fang} Fang, M. H.; Pham, H. M.; Qian, B.; Liu, T. J.; Vehstedt, E. K.; Liu, Y.; Spinu, L.; Mao, Z. Q. Superconductivity Close to Magnetic Instability in $\textrm{Fe(Se}_{1-x}Te_{x})_{0.82}$. {\it Phys. Rev.} B {\bf 2008}, {\it 78}, 224503-5

\bibitem {yeh} Yeh, K. W.; Ke, C. T.; Huang, T. W.; Chen, T. K.; Huang, Y. L.; Wu, P. M.; Wu, M. K. Tellurium Substitution Effect on Superconductivity of the $\alpha$-Phase Iron Selenide. {\it Europhys. Lett.} {\bf 2008}, {\it 84}, 37002-4

\bibitem{yeh1} Yeh, K. W.; Ke, C. T.; Huang, T. W.; Chen, T. K.; Huang, Y. L.; Wu, P. M.; Wu, M. K. Superconducting $\textrm{Fe}\textrm{Se}_{1-x}\textrm{Te}_{x}$ Crystals Grown by Optical Zone Melting Technique. {\it Cryst. Growth Des.} {\bf 2009}, {\it 9}, 4847-4851

\bibitem{gresty} Gresty, N. C.; Takabayashi, Y.; Ganin, A. Y.; McDonald, M. T.; Claridge, J. B.; Giap, D.; Mizuguchi, Y.; Takano, Y.; Kagayama, T.; Ohishi, Y. Structural Phase Transitions and Superconductivity in $\textrm{Fe}_{1+\delta}\textrm{Se}_{0.57}\textrm{Te}_{0.43}$ at Ambient and Elevated Pressures. {\it J. Am. Chem. Soc.} {\bf 2009}, {\it 131}, 16944-16952

\bibitem{grechnev} Grechnev, G. E.; Panfilov, A. S.; Fedorchenko, A. V.; Desnenko, V. A.; Gnatchenko, S. L.; Tsurkan, V.; Deisenhofer, J.; Loidl, A.; Chareev, D. A.; Volkova, O. S.; et al. Magnetic Properties of Novel FeSe(Te) Superconductors. {\it J. Magn. Magn. Mater.} {\bf 2012}, {\it 324}, 3460-3463

\bibitem {viennois1} Viennois, R.; Giannini, R.; van der Marel, D.; \v{C}ern\'y, R. Effect of Fe Excess on Structural, Magnetic and Superconducting Properties of  Single-Crystalline $\textrm{Fe}_{1+x}\textrm{Te}_{1-y}\textrm{Se}_y$. 
  {\it J. Solid State Chem.} {\bf 2010}, {\it 183}, 769-775
  
  \bibitem{margadonna} Margadonna, S.; Takabayashi, Y.; McDonald, M. T.; Kasperkiewicz, K.; Mizuguchi, Y.; Takano, Y.; Fitch, A. N.; Suarde, E.; Prassides, K. Crystal Structure of the New $\textrm{FeSe}_{1-x}$ Superconductor. {\it Chem. Commun.} {\bf 2008}, {\it 7345}, 5607-5609

\bibitem{mcqueen} McQueen, T. M.; Williams, A. J.; Stephens, P. W.; Tao, J.; Zhu, Y.; Ksenofontov, V.; Casper, F.;  Felser, C.; Cava, R. J. Tetragonal-to-Orthorhombic Structural Phase Transition at 90 K in the Superconductor $\textrm{Fe}_{1.01}\textrm{Se}$. {\it Phys. Rev. Lett.} {\bf 2009}, {\it 103}, 057002-7
 
\bibitem {kurmaev1} Kurmaev, E. Z.; McLeod, J. A.; Buling, A.; Skorikov, N. A.; Moewes, A.; Neumann, M.; Korotin, M. A.; Izyumov, Y. A.; Ni, N.; Canfield, P. C. Contribution of Fe $3d$ States to the Fermi Level of $\textrm{Ca}\textrm{Fe}_{2}\textrm{As}_2$. {\it Phys. Rev.} B {\bf 2009}, {\it 80}, 054508-6

\bibitem {kurmaev2} Kurmaev, E. Z.; McLeod, J. A.; Skorikov, N. A.; Finkelstein, L. D.; Moewes, A.; Izyumov, Y. A.;  Clarke, S.  Identifying Valence Structure in LiFeAs and NaFeAs with Core-Level Spectroscopy.
{\it J. Phys.: Condens. Matter} {\bf 2009}, {\it 21}, 345701-6

\bibitem{mcleod5} McLeod, J. A.; Buling, A.; Green, R. J.; Boyko, T. D.; Skorikov, N. A.; Kurmaev, E. Z.; Neumann, M.; Finkelstein, L. D.; Ni, N.; Thaler, A.; Bud'ko, S. L.; et al. Effect of $3d$-Doping on the Electronic Structure of $\textrm{Ba}\textrm{Fe}_2\textrm{As}_2$. {\it J. Phys.: Condens. Matter} {\bf 2012}, {\it 24},  215501-11

 \bibitem {singh} Singh, D. J.; Du M. H. Density Functional Study of $\textrm{LaFeAsO}_{1-x}\textrm{F}_{x}$: A Low Carrier Density Superconductor Near Itinerant Magnetism. {\it Phys. Rev. Lett.} {\bf 2008}, {\it 100}, 237003-4
 
 \bibitem {singh1} Singh, D. J. Electronic Structure and Doping in $\textrm{BaFe}_{2}\textrm{As}_{2}$ and LiFeAs: Density Functional Calculations. {\it Phys. Rev.} B {\bf 2008}, {\it 78}, 094511-7

\bibitem{aichhorn1} Aichhorn, M.; Pourovskii, L.; Vildosola, V.; Ferrero, M.; Parcollet, O.; Miyake, T.; Georges, A.; Biermann, S. Dynamical Mean-Field theory Within an Augmented Plane-wave Framework: Assessing Electronic Correlations in the Iron Pnictide LaFeAsO. {\it Phys. Rev.} B {\bf 2009}, {\it 80}, 085101-15

 \bibitem{yang1} Yang, W. L.; Sorini, A. P.; Chen, C-C.; Moritz, B.; Lee, W.-S.; Vernay, F.; Olalde-Velasco, P.;  Denlinger, J. D.; Delley, B.; Chu, J.-H.; et al. Evidence for Weak Electronic Correlations in Iron Pnictides.
{\it Phys. Rev.} B {\bf 2009}, {\it 80}, 014508-10

\bibitem {subedi} Subedi, A.; Zhang, L.; Singh, D. J.; Du, M. H. Density Functional Study of FeS, FeSe, and FeTe: Electronic Structure, Magnetism, Phonons, and Superconductivity.  {\it Phys. Rev.} B {\bf 2008}, {\it 78}, 134514-6

\bibitem{tamai} Tamai, A.; Ganin, A. Y.; Rozbicki, E.; Bacsa, J.; Meevasana, W.; King, P. D. C.; Caffio, M.; Schaub, R.; Margadonna, S.; Prassides, K.; et al. Strong Electron Correlations in the Normal State of the Iron-Based $\textrm{Fe}\textrm{Se}_{0.42}\textrm{Te}_{0.58}$ Superconductor Observed by Angle-Resolved Photoemission Spectroscopy.  {\it Phys. Rev. Lett.} {\bf 2010}, {\it 104}, 097002-4

\bibitem{homes} Homes, C. C.; Akrap, A.; Wen, J. S.; Xu, Z. J., Lin, Z. W.; Li, Q.; Gu, G. D. Electronic Correlations and Unusual Superconducting Response in the Optical Properties of the Iron Chalcogenide $\textrm{Fe}\textrm{Te}_{0.55}\textrm{Se}_{0.45}$. {\it Phys. Rev.} B {\bf 2010}, {\it 81}, 180508(R)-4

\bibitem{aichhorn} Aichhorn, M.; Biermann, S.; Miyake, T.; Georges, A.; Imada, M. Theoretical Evidence for Strong Correlations and Incoherent Metallic State in FeSe. {\it Phys. Rev.} B {\bf 2010}, {\it 82}, 064504-5

\bibitem{pourret} Pourret, A.; Malone, L.; Antunes, A. B.; Yadav, C. S.; Paulose, P. L.; Fauqu\'e, B.; Behnia, K. Strong Correlation and Low Carrier Density in $\textrm{Fe}_{1+y}\textrm{Te}_{0.6}\textrm{Se}_{0.4}$ as Seen From its Thermoelectric Response. {\it Phys. Rev.} B {\bf 2011}, {\it 83}, 020504(R)-4

\bibitem{iperez14a} Perez, I.; McLeod, J. A.; Green, R. J.; Escamilla, R.; Ortiz, V.; Moewes, A. Electronic Dtructure of Co-substituted FeSe Superconductor Probed by Soft X-ray Spectroscopy and Density Functional Theory. {\it Phys. Rev.} B {\bf 2014}, {\it 90}, 014510-8
 
  \bibitem{gomez} G\'omez, R. W.; Marquina, V.; P\'erez-Mazariego, J. L.; Escamilla, R.; Escudero, R.; Quintana, M.;  Hern\'andez-G\'omez, J. J.; Ridaura, R.; Marquina, M. L. Effects of Substituting Se with Te in the FeSe Compound: Structural, Magnetization and M\"ossbauer Studies; cond-mat: arxiv: 0910.2504.
 
 \bibitem{mizuguchi} Mizuguchi, Y.; Tomioka, F.; Tsuda, S.; Yamaguchi, T.; Takano, Y. Substitution Effects on FeSe Superconductor. {\it J. Phys. Soc. Jpn.} {\bf 2009}, {\it 78}, 074712-5

\bibitem {jia} Jia, J. J.; Callcott, T. A.; Yurkas, J.; Ellis, A. W.; Himpsel, F. J. First Experimental Results from IBM/TENN/TULANE/LLNL/LBL Undulator Beamline at the Advanced Light Source. {\it Rev. Sci. Instrum.} {\bf 1995}, {\it 66}, 1394-1397

\bibitem {blaha} Blaha, P.; Schwarz, K.; Madsen, G. K. H.; Kvasnicka, D.; Luitz, J. WIEN2k An Augmented Plane Wave + Local Orbitals Program for Calculating Crystal Properties; Schwarz, K. Ed.; Techn. Universit\"at Wien, Vienna Austria, {\bf 2001}; ISBN 3-9501031-1-2.

\bibitem {perdew} Perdew, J. P.; Burke, K.; Ernzerhof, M. Generalized Gradient Approximation Made Simple. {\it Phys. Rev. Lett.} {\bf 1996}, {\it 77}, 3865-3868

\bibitem {schwarz} Schwarz, K.; Neckel, A.; Nordgren, J. On the X-ray Emission Spectra of FeAl. {\it J. Phys. F: Met. Phys.} {\bf 1979}, {\it 9}, 2509-2521

\bibitem {kurmaev3} Kurmaev, E. Z.; Ankudinov, A. L.; Rehr, J. J.; Finkelstein, L. D.; Karimov, P. F.; Moewes, A. The L2:L3 Intensity Ratio in Soft X-ray Emission Spectra of $3d$-Metals. {\it J. Electron. Spectrosc. Relat. Phenom.} {\bf 2005}, {\it 148}, 1-4

\bibitem {gao} Gao, X.; Qi, X.; Tan, S. C.; Wee, A. T. S.; Yu, X.; Moser, H. O. Thickness Dependence of X-ray Absorption and Photoemission in Fe Thin Films on Si(0 0 1).  {\it J. Electron. Spectrosc. Relat. Phenom.} {\bf 2006}, {\it 151}, 199-203

\bibitem {prince} Prince, K. C.; Matteucci, M.; Kuepper, K.; Chiuzbaian, S. G.; Bartkowski, S.; Neumann, M. Core-level Spectroscopic Study of FeO and $\textrm{FeS}_2$. {\it Phys. Rev.} B  {\bf 2005}, {\it 71}, 085102-9

\bibitem {thole} Thole, B. T.; van der Laan, F. Branching Ratio in X-ray Absorption Spectroscopy.  {\it Phys. Rev.} B {\bf 1988}, {\it 38}, 3158-3171

\bibitem{yablonskikh} Yablonskikh, M. V.; Yarmoshenko. Y. M.; Grebennikov, V. I.; Kurmaev, E. Z.; Butorin, S. M.; Duda, L. C.; Nordgren, J.; Plogmann, S.; Neumann, M. Origin of Magnetic Circular Dichroism in Soft X-ray Fluorescence of Heusler Alloys at Threshold Excitation. {\it Phys. Rev.} B {\bf 2001}, {\it 63}, 235117-10

\bibitem{wu3} Wu, W. Modelling the Electronic Structure and Magnetic Properties of LiFeAs and FeSe Using Hybrid-Exchange Density Functional Theory. {\it Solid State Commun.} {\bf 2013}, {\it 161}, 23-28

\bibitem{gnezdilov} Gnezdilov, V. Pashkevich, Y. G.; Lemmens, P.; Wulferding, D.; Shevtsova, T.; Gusev, A.; Chareev, D.; Vasiliev, A. Interplay Between Lattice and Spin States Degree of Freedom in the FeSe Superconductor: Dynamic Spin State Instabilities. {\it Phys. Rev.} B {\bf 2013}, {\it 87}, 144508-8

\bibitem{gnezdilov1} Gnezdilov, V.; Pashkevich, Y.; Lemmens, P.; Gusev, A.; Lamonova, K.; Shevtsova, T.; Vitebskiy, I.; Afanasiev, O.; Gnatchenko, S.; Tsurkan, V.; et al. Anomalous Optical Phonons in FeTe Chalcogenides: Spin State, Magnetic Order, and Lattice Anharmonicity. {\it Phys. Rev.} B {\bf 2011}, {\it 83}, 245127-6

\bibitem{medici} de Medici, L.; Mravlje, J.; Georges, A. Janus-Faced Influence of Hund's Rule Coupling in Strongly Correlated Materials. {\it Phys. Rev. Lett.} {\bf 2011}, {\it 107}, 256401-4

\bibitem{haule} Haule, K.; Kotliar, G. Coherence-Incoherence Crossover in the Normal State of Iron Oxypnictides and Importance of Hund's Rule Coupling. {\it New J. Phys.} {\bf 2009}, {\it 11}, 025021-14

\bibitem{johannes} Johannes, M. D.; Mazin, I. I. Microscopic Origin of Magnetism and Magnetic Interactions in Ferropnictides. {\it Phys. Rev.} B {\bf 2009}, {\it 79}, 220510(R)-4

\bibitem{kawai1} Kawai, J.; Maeda, K.; Nakajima, K.; Gohshi, Y. Relation Between the Copper $L$ X-ray Fluorescence and $2p$ X-ray Photoelectron Spectroscopies. {Phys. Rev.} B {\bf 1993}, {\it 48}, 8560-8566

\bibitem{kawai2} Kawai, J. Intesity Ratio of the Transition-Metal $L_{\alpha}$ and $L_{\beta}$ Lines. {\it Rigaku Journal} {\bf 2001}, {\it 18}, 31-37

\bibitem{laan} van der Laan, G.; Kirkman, I. W. The 2p Absorption Spectra of 3d Transitions Metal Compounds  in Tetrahedral and Octahedral Symmetry. {\it J. Phys.: Condens. Matter} {\bf 1992}, {\it 4}, 4189-4204

 
\end{thebibliography}

\end{document}